# Efficient phase-only wavefront shaping at 250 kHz for multimode fiber endomicroscopy

**Pau Riera Rodríguez,[1,*] Antonio M. Caravaca-Aguirre,[2] Omer Tzang,[2] Alexander Payne,[3] Stephen Hamann,[3] and Rafael Piestun[1,*]**

[1]*Department of Electrical, Computer, and Energy Engineering, UCB 425, University of Colorado Boulder, Boulder, CO 80309, USA*

[2]*Modendo Inc.,1898 S Flatiron Ct #115, Boulder, CO 80301*

[3]*Silicon Light Machines, San Jose, California, United States*

*pau.rierarodriguez@colorado.edu, rafael.piestun@colorado.edu

## Abstract

Wavefront shaping through multimode fibers (MMFs) enables diffraction-limited fluorescence imaging with hair-thin probes, at a speed and light budget bounded by the wavefront modulator. Existing modulators provide at most two of three key characteristics: high refresh rate, (high) modal dimensionality, and light efficiency. Here, we demonstrate wavefront control through an MMF using a tailored phase-only microelectromechanical modulator that combines all three, with 8,192 elements arranged in a 32 × 256 array, 7-bit phase control, and a 250 kHz full-frame refresh rate. Direct on-axis phase modulation delivers $70 \pm 5\%$ of the incident light into the modulated field and eliminates both off-axis encoding optical losses and hologram computation. The generated foci at the distal end of the fiber remain isotropic with high fidelity. We use singular-value decomposition analysis of the measured transmission to identify the effective degrees of freedom of our system. By scanning the focus at multiple depths, we demonstrate volumetric fluorescence imaging. These results open a new operating regime for adaptive in vivo multimode-fiber endoscopy.

## Introduction

Multimode fibers (MMFs) support thousands of spatial modes through a core with a diameter comparable to the width of a human hair [1]. Interference and coupling among these modes scramble the transmitted optical field into a speckle pattern, preventing direct image transmission [2]. Despite the apparent randomness, propagation remains linear and deterministic, which can be described by a transmission matrix (TM), enabling direct mapping between the input and output fields after propagation through the complex media [3-5]. Wavefront shaping techniques can compensate for this transformation, generate a focus, and transmit an image at or beyond the distal facet [6]. More broadly, wavefront shaping enables precise control over the amplitude, phase, and polarization of light [7-14], as well as its temporal and spectral properties [15-19]. Consequently, this multidimensional control can be harnessed to either, enhance or suppress nonlinear optical processes through highly complex media [2]. Beyond overcoming scattering, recent advances have further demonstrated that speckle fields themselves can be engineered to exhibit deterministic propagating dynamics and statistical properties, providing additional degrees of freedom for complex light manipulation [21-23].

Among the applications of wavefront shaping, MMF endomicroscopy is particularly attractive for minimally invasive imaging. By scanning a focus at the distal end, a simple MMF can function as a compact imaging probe without distal scanning optics [24, 25]. Advances in wavefront control have enabled a broad range of imaging modalities through MMF, including fluorescence [26-29], coherent [30, 31], nonlinear [32-37], and photoacoustic [38-42]. In vivo deep-brain fluorescence imaging through an MMF was first reported in 2017 [43], followed by detailed demonstrations of neural structure and activity [44-48]. These applications require rapid scanning, large number of control modes to achieve high signal enhancement, and sufficient optical power delivery at the sample.

Therefore, meeting MMF wavefront shaping requirements calls for a wavefront modulation approach that combines a high refresh rate, high modal dimensionality, and efficient delivery of laser power. Liquid-crystal Spatial Light Modulators (LC-SLMs) provide many spatial degrees of freedom but typically operate at tens to a few hundred hertz [49]. Digital Micromirror Devices (DMDs) based on microelectromechanical systems (MEMS) technology were introduced to wavefront shaping to overcome the speed limit [50], operating at ~23 kHz, and enabled the first real-time focusing through a bending MMF [51]. However, it requires binary off-axis holographic encoding, which reduces the optical power available in the desired diffraction order and introduces additional computational steps. Similarly, wavefront shaping at 350 kHz was previously demonstrated using a one-dimensional Grating Light Valve (GLV), although the limited number of controllable spatial modes and fill factor, limits its implementation in high-fidelity MMF imaging [52]. More recently, efficient wavefront control was enabled by using a two-dimensional MEMS phase-only modulator, Phase Light Modulator (PLM), operating at 1.44 kHz [53]. These developments underscore the critical need for faster modulators that retain optical efficiency and high modal control [54, 55].

The effective number of independently controllable channels is bounded by the fiber's finite modal dimensionality per polarization [1]. Increasing the SLM dimensionality and input modes beyond this bound does not necessarily increase the number of independently accessible fiber modes and often leads to oversampling and increased computational burden. Megapixel devices are needed for off-axis holographic encoding as each controlled mode is mapped across several pixels, whereas direct phase control maps each pixel to a different mode, avoiding encoding overhead. A modulator with thousands rather than millions of directly controlled phase elements can therefore offer a fast and efficient architecture suited to MMF wavefront shaping.

In this work, we demonstrate high-speed wavefront shaping and imaging through an MMF using a tailored two-dimensional Displacement Phase Modulator (DPM) [56]. The DPM consists of a 32 x 256 array of 8192 reflective MEMS pixels, enabling direct on-axis phase modulation at a full-frame refresh rate of 250 kHz, corresponding to 4 μs per full displayed pattern. We use 4,096 input pixels (modes) for wavefront control and the remaining half of the array as an internal reference for TM measurement. Therefore, we eliminate the additional encoding of off-axis and reduce the computational cost of generating focusing masks. The generated foci remain isotropic across the output facet despite the DPM's anisotropic geometry. We further use singular value decomposition to characterize the measured TM [5, 57], revealing a dominant singular channel and a remaining normalized spectrum with a decay similar to that obtained with off-axis modulation. A Marchenko-Pastur (MP) calibrated noise threshold provides an estimate of the effective (experimentally resolved) degrees of freedom [58, 59]. After numerical subtraction of the unmodulated contribution from the focusing measurements, the enhancement is consistent with the phase-only theoretical prediction [60]. Finally, we demonstrate two-dimensional fluorescence imaging of fixed biological cells and volumetric imaging of microspheres suspended in agar gel. Axial multi-plane calibration enables three-dimensional imaging without moving the sample, reducing fiber perturbations. We perform shift variant deblurring to improve image fidelity and remove background. These experiments demonstrate, to our knowledge, the highest pattern rate (250 kHz) direct 2D phase modulation wavefront shaping, and the fastest point-scanning fluorescence imaging through an MMF.

Figure 1 shows the principle of wavefront shaping and imaging through an MMF at 250 kHz. A 2D MEMS phase SLM displays an optimized phase mask to generate a focus at the distal tip of the MMF. The excited fluorescence is acquired through the same MMF and collected by a photomultiplier tube (PMT). Half of the array shapes the wavefront shaping field while half launches an internal reference through the same fiber during transmission-matrix measurement. Because each mask implements a phase assignment rather than an encoded diffractive pattern, the mask synthesis requires one order of magnitude fewer operations than off-axis encoding.

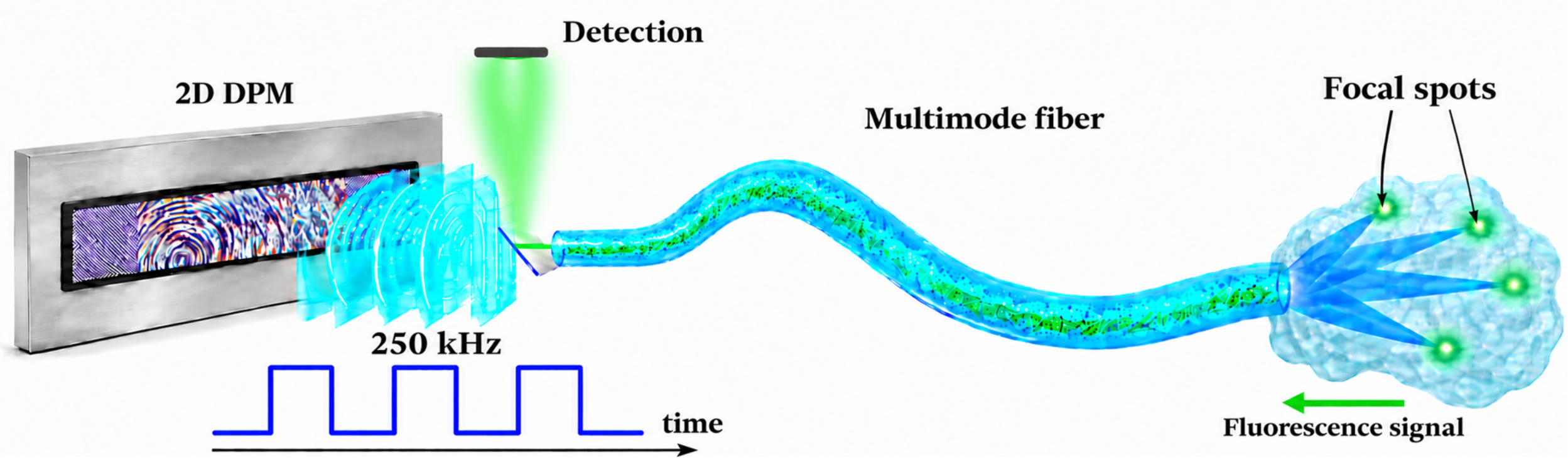

**Fig.1. Multimode fiber endomicroscopy at 250 kHz.** A two-dimensional microelectromechanical phase-only modulator operating at 250 kHz shapes the input wavefronts into a multimode fiber to form foci at the distal facet and at calibrated planes beyond it. The excited fluorescence by each focus is collected back through the same fiber and recorded by a photomultiplier tube (PMT), enabling volumetric imaging with high resolution.

## Results

### Wavefront shaping performance: efficiency, speed and computational cost

Fig. 2 a compares the DPM with other common SLMs used for wavefront shaping in terms of optical efficiency and full-frame refresh rate. We define the modulator efficiency as

$$\epsilon_{opt} = \frac{P_{\text{field}}}{P_{\text{incident}}} \tag{1}$$

where $P_{\text{field}}$ is the optical power in the modulated field and $P_{\text{incident}}$ is the power incident on the modulator. This definition excludes optical losses from fiber coupling and the fraction of power transmitted into the generated focus, because they are common to every device.

For the operating conditions plotted in Fig. 2a, the efficiencies are approximately 85±5% for LC-SLMs, 82.5±2.5% for the PLM, 4 ±4% for DMDs, and 70±5% for the DPM [56]. LC-SLMs typically operate at tens to hundreds of hertz [49], the PLM operates at 1.44 kHz [53], and DMDs reach 22.7 kHz [50]. The DPM combines these two key characteristics with high optical efficiency and a full-frame refresh rate of 250 kHz [56]. Overall, to assess the combined efficiency and speed of the system, we define the figure of merit

$$\mathcal{M} = \epsilon_{opt} \cdot f_r \tag{2}$$

where $f_r$ is the modulator refresh rate. This index expresses the pattern rate weighted by optical efficiency. For the DPM, $\mathcal{M} = 1.75 \times 10^5\ s^{-1}$, compared with approximately $1.19 \times 10^3\ s^{-1}$ for the PLM, $9.8 \times 10^2\ s^{-1}$ for the DMD, and $8.5 - 85\ s^{-1}$ for the LC-SLM. $\mathcal{M}$ compares devices at a comparable number of controlled channels, namely a device with many more channels trades update rate for control dimensionality, which $\mathcal{M}$ alone does not capture.

The refresh rate also sets a lower bound on TM acquisition time. For three-step phase-shifting interferometry with $N$ controlled input modes, calibration requires $3N$, yielding an ideal modulator-limited acquisition time of

$$t_{TM} = \frac{3N}{f_r} \tag{3}$$

For $N = 4{,}096$ input modes used here, this corresponds to 49 ms at 250 kHz, 0.54 s at 22.7 kHz, 8.5 s at 1.44 kHz, and 2.05−20.5 min at 10-100 Hz. These values assume that the detector has the bandwidth to record every displayed pattern. Furthermore, TM acquisition can be performed with an external reference, requiring only one phase step per mode but yielding higher phase uncertainties when compared with our common path reference. The metrics of Fig. 2a assume the similar controllable modal dimensionality required for MMF as discussed below.

After TM measurements and basis conversion, each phase mask is obtained by phase conjugation and converted into a modulation pattern using the encoding scheme required by the device. For $l$ phase masks, each corresponding to a focus spot, we estimate the required number of operations, $O$ as

$$O = lNqs \tag{4}$$

where $s$ is the number of pixels per controlled input mode, and $q$ denotes the encoding overhead per pixel, expressed as an unweighted operation count. Fig. 2 b compares the number of operations per focus ($l = 1$) from 256 to 8,192 controlled input modes. We assume one pixel per input mode for direct phase modulation ($s = 1$) and eight pixels for both off-axis implementations ($s = 8$).

These assumptions yield $N$, $16N$, and $24N$ operations per focus for direct phase, off-axis phase, and binary amplitude Lee encoding, respectively. Therefore, direct-phase modulation reduces the operation count by factors of 16 and 24 relative to off-axis and binary-amplitude encoding, respectively. For a fixed number of controllable modes, the operation count scales with the required steps for holographic encoding, with each step having similar computational complexity. Direct-phase modulation also diminishes the delay between TM acquisition and focusing, which could facilitate recalibration when imaging through highly dynamic media, such as biological tissue, or MMF bending compensation. Altogether, the low pixel count, matched to the typical number of input modes required for MMFs with the higher refresh rate, motivates the use of DPMs for scalable and fast wavefront shaping systems.

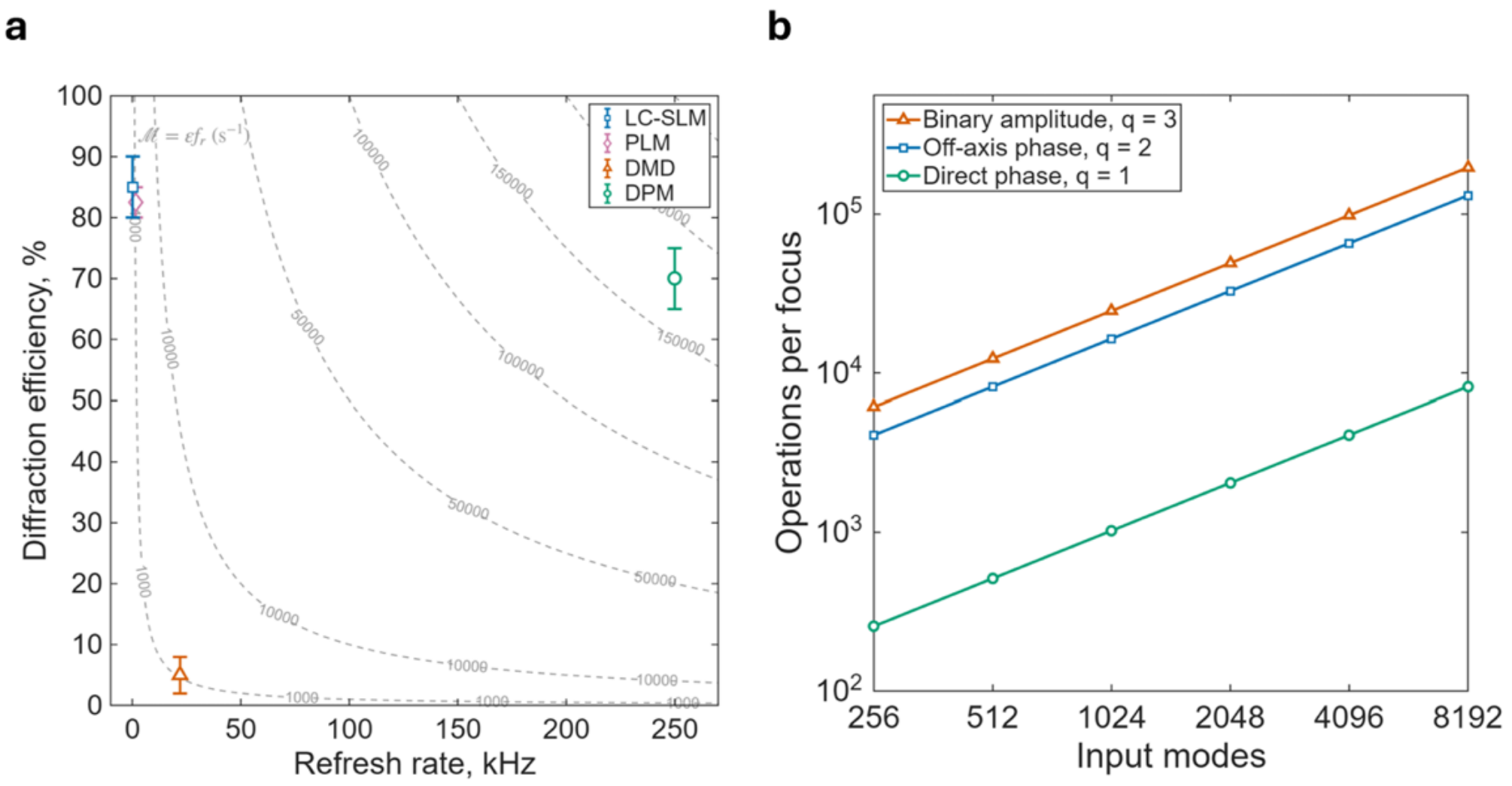

**Fig. 2. Diffraction efficiency vs refresh rate, and comparison of operations per focus across wavefront modulation techniques. a** Diffraction efficiency $\epsilon_{opt}$ as a function of refresh rate $f_r$ in kilohertz (kHz), and figure of merit $\mathcal{M} = \epsilon_{opt} \cdot f_r$, mapped as contours, for a given control dimensionality. Results are compared for the liquid crystal SLM (LC-SLM), phase light modulator (PLM), digital micromirror device (DMD), and displacement phase modulator (DPM). **b.** Estimated operations required to calculate and generate one focusing pattern as a function of the number of input modes for three different wavefront modulation techniques.

## Performance analysis

The experimental setup is presented in Methods. Before calibration, the distal output of the MMF exhibits a random speckle pattern (Fig. 3 a). Despite the anisotropic geometry of the DPM, the output is consistent with fully developed speckle statistics. We measure a TM $\in \mathbb{C}^{m \times n}$, where $m = 25{,}600$ is the number of camera pixels and $n = 4{,}096$ is the number of controlled input modes, using a Hadamard input basis and three-step phase-shifting interferometry with an internal reference of 4,096 pixels. Once the TM is measured, phase conjugation of the corresponding row generates a focus at a selected output location on the distal facet of the MMF (Fig. 3 b), and a superposition of focus spots (Fig. 3 c).

To characterize the focusing temporal response, we alternate between an optimized focusing pattern and a flat phase pattern on the DPM while recording the intensity with an avalanche photodiode (APD). This allows direct measurement of the switching dynamics between the focused and unfocused (speckle) states. Fig. 3 d shows the measured temporal response of the switching transition, confirming wavefront modulation at the full update rate of 250 kHz, corresponding to a pattern update period of 4 μs.

We quantify focusing fidelity using the enhancement, defined as the ratio of the peak focal intensity to the mean residual background intensity. For each focus, we use a dual-exposure measurement to combine the unsaturated focal peak intensity recorded at a short exposure with the background recorded at a longer exposure. The background is averaged over the entire field of view, excluding the masked region surrounding the focus. The enhancement is expressed as

$$\eta = \frac{I_n}{\langle I \rangle} \quad (5)$$

where $I_n$ denotes the $n$-th exposure-corrected peak intensity of the generated focal spot and $\langle I \rangle$ is the average intensity of the residual speckle background. Across the measured foci, the mean enhancement is ~1,400, with individual foci reaching values as high as 2,700 (Fig. 3 e).

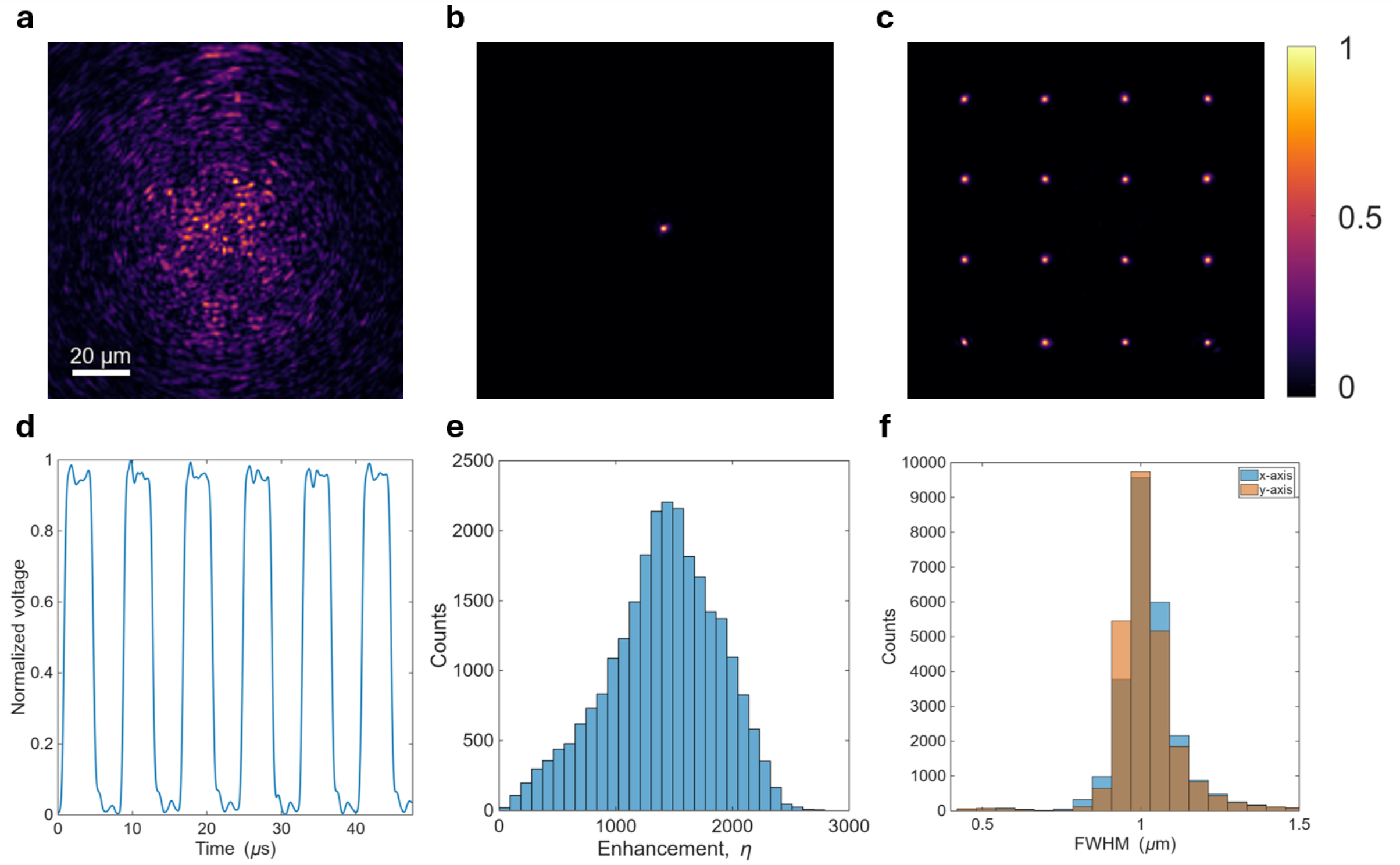

**Fig. 3. High-speed operation and focusing fidelity. a** Speckle pattern at the distal facet of the multimode fiber before calibration. **b** Focus generated after optimization. **c** Superposition of 16 equally spaced foci across the facet. The colorbar represents intensity in arbitrary units. **d** Normalized output voltage from an avalanche photodiode monitoring the intensity between an optimized mask and a flat phase mask, demonstrating a pattern update period of 4 μs. **e** Histogram of enhancement across 25,600 measured foci, with a mean enhancement of 1,400 and a maximum of 2,700. **f** Histogram of the measured Full Width Half Maximum (FWHM) along the horizontal (X) and vertical (Y) directions, with median values of 1.028 μm and 1.022 μm, respectively.

We further assess spatial isotropy by measuring the full width at half maximum (FWHM) along both lateral axes. The horizontal and vertical axes closely overlap, with median values of 1.028 µm and 1.022 µm, respectively (Fig. 3 f). The nearly equal widths show that the foci remain approximately isotropic despite the 1:8 aspect ratio of the DPM, consistent with the redistribution of the input field through mode coupling within the MMF.

## Effective number of degrees of freedom

In order to investigate the effective number of degrees of freedom and to examine the influence of the unmodulated contribution, we analyze the singular value decomposition (SVD) of the measured TM. The singular value spectrum exhibits a dominant first singular value, indicating a strong common or low-rank contribution, consistent with the unmodulated field retained in the on-axis configuration (Fig. 4). The first singular channel accounts for approximately 22% of the squared Frobenius norm of the measured TM. The spectrum alone is not yet enough to resolve the effective number of degrees of freedom (resolved dimensionality), $N_{eff}$, so we require the noise level of the TM

$$N_{eff} = \{n: s_n > \tau\} \quad (6)$$

where $\tau$ is the noise-calibrated singular value threshold which we estimate by fitting the spectral tail to the MP model and apply the Gavish-Donoho optimal threshold [59], as detailed in Methods. This procedure gives $N_{eff} = 2{,}649$; varying the estimated noise level by $\pm 5\%$ changes the retained count by $+58/-53$. This is an estimate of the accessible dimensionality of the measured TM, rather than the total number of physical modes supported by the fiber. Random-matrix theory is used only for the measurement noise, not for the fiber.

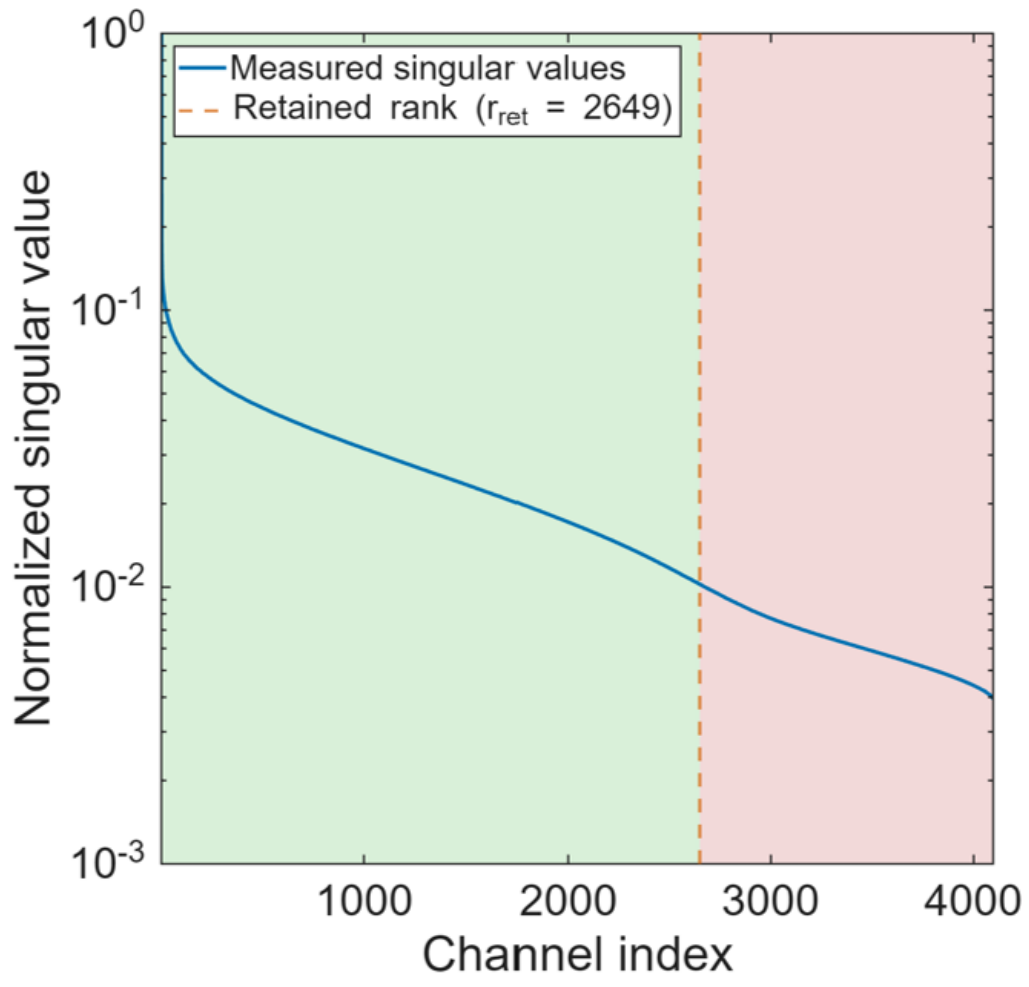

**Fig. 4. Transmission eigenchannel spectra and effective degrees of freedom.** Normalized singular value spectra of the measured transmission matrix versus the eigenchannel index. The dashed line marks the estimated retained rank, with the green shading indicating the effective degrees of freedom, and red the channels attributed to noise.

For $N_{eff} = 2{,}649$, the estimate is 2,100. The measured mean enhancement of 1,400 includes the unmodulated background and, after estimated numerical subtraction of this contribution, the recorded focus frames raise the mean enhancement to approximately 1,800, approaching the theoretical case. The remaining difference is consistent with phase errors from the slight curvature of the pistons [56]. We note that the SVD estimate does not reflect the unmodulated contribution in the recorded focus frames.

**Fluorescence imaging**

We next exploit the high refresh rate of the DPM for fluorescence imaging through the MMF. Following the TM calibration at 200 µm beyond the distal fiber facet, we sequentially scan and evaluate the focusing performance across the facet. The emitted fluorescence is collected through the same fiber, detected by a PMT, and digitized by a high-speed data acquisition (DAQ) synchronized with the mask sequence. At a pattern period of 4 µs, displaying 25,600 masks requires 102.4 ms per frame. An image is reconstructed by assigning each measured fluorescence signal to its corresponding focal position.

We first image fixed bovine pulmonary artery endothelial (BPAE) cells (Fig. 5 a-c) with a field of view of 69 x 69 $\mu m^2$. Each fluorescent measurement contains contributions from both the out-of-focus and residual speckle background. To improve image contrast and suppress background contributions, we apply spatially variant Richardson-Lucy deblurring using the experimentally measured high-dynamic range point-spread functions (PSFs) [61, 62]. The reconstruction is described in Methods.

For volumetric imaging, we calibrate TMs at six axial planes separated by 10 µm, spanning a total axial range of 50 µm by translating the calibration module along the propagation direction with a piezo stage. We then insert the MMF into agar gel containing fluorescent microspheres and reconstruct fluorescence images at calibrated axial positions of z=0, 20, and 40 µm from the distal fiber tip (Fig. 5 d-f). Because the TMs are calibrated in air, propagation in the sample with an index of refraction of 1.33 shifts the focal depths, yielding nominal depths of 0, 27, and 53 µm in the sample. Once calibrated, the high refresh rate of the DPM enables rapid lateral scanning and between axial planes without mechanically translating the sample during imaging, while simultaneously minimizing any perturbations. The fluorescence distribution is sampled throughout the calibrated volume and reconstructed in three dimensions.

The resulting axially measured foci exhibit near-circular symmetry. We evaluate the FWHM along the horizontal (X) and vertical (Y) directions of the generated foci over the calibrated range, as shown in Fig. 5 g. The enhancement boosts after free-space propagation arise from additional interference and remains constant at mean values of 1,400, reaching maxima of nearly 3,000 (Fig. 5 h). These results indicate that the strong mode mixing within the MMF effectively scrambles the anisotropy introduced by the geometry of the DPM.

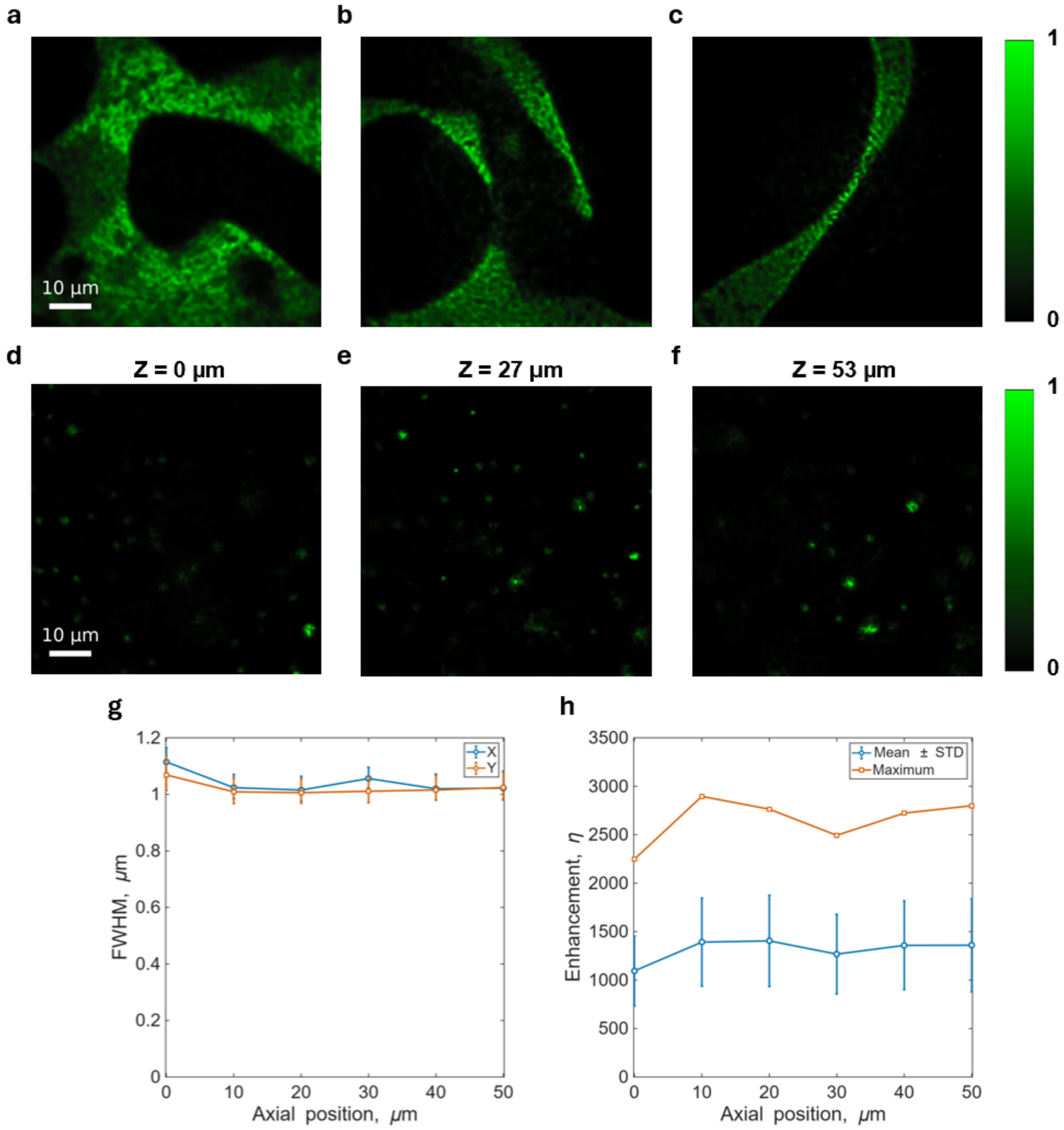

**Fig. 5. High-resolution fluorescence imaging. a-c** Fluorescence images of different regions of fixed BPAE cells acquired through the MMF, shown after shift-variant deblurring. The TM is measured at 200 µm away from the facet. **d-f** Fluorescence bead reconstructions at different axial locations, 0, 27, and 53 µm, respectively. The colorbar represents normalized fluorescence intensity in arbitrary units. All images have been reconstructed after 30 deblurring iterations. **g** Lateral Full Width at Half Maximum (FWHM) of the generated focus spots. **h** Axial enhancement for six independent TM calibrations at distances 0 through 50 µm from the distal end of the fiber.

## Discussion

In this work, we demonstrate wavefront shaping at 250 kHz and fluorescence imaging through an MMF using direct phase modulation of thousands of controlled channels in a MEMS phase modulator. With a 4 µs refresh period, the system enables wavefront shaping eleven times faster than DMD-based wavefront shaping at 22.7 kHz, while maintaining high fidelity with 4,096 directly controlled channels. Despite the anisotropic geometry of the DPM, the generated focal spots are isotropic due to azimuthal mixing among near-degenerate modes, showing that anisotropic modulator arrays can serve MMFs without loss of resolution. The mean enhancement of 1,400 rises to at least 1,800 once the unmodulated light is accounted for, approaching the value of about 2,100 expected for phase conjugation over the effective rank of about 2,600. Volumetric imaging needs no moving parts since any calibrated plane is addressed in a 4 μs period; this is critical for endomicroscopy where there is no sample stage, and depth must be selected by the wavefront.

For fluorescence imaging, the efficiency of on-axis modulation comes with a cost: the unmodulated light that reduces enhancement and contributes to background fluorescence. This background is part of the measured focal intensity distribution at each scan position, which forms the forward model of the shift-variant deblurring. Fortunately, unmodulated light is not fundamental; most of it can be removed by spatial filtering or, in future devices, by patterning the perimeter pixels with a fixed grating, so that their reflected light is diffracted outside the aperture.

Although MMFs can support thousands of guided modes, a practical wavefront-shaping system typically accesses only a subset of effective transmission eigenchannels (14,184 fiber modes per polarization, 4,096 controlled channels, about 2,600 effective degrees of freedom in our case). The fiber modal capacity sets the physical upper bound on the rank of the measured TM, while the optical system determines which portion of this space is excited and detected. Increasing the number of controllable input degrees of freedom can improve focusing fidelity, but additional input channels beyond the number of fiber modes do not provide independent eigenchannels and can increase calibration and computational requirements without proportional gains in enhancement. The singular value spectrum is more informative than input channel counting alone, as it characterizes the dimensionality and transmission power of the measured eigenchannels. Thus, the SVD analysis complements the enhancement measurements by characterizing the effective transmission-channel structure beyond the number of controlled inputs. The measured spectrum shows a dominant first singular value, while the remaining spectrum displays a decay similar to that obtained with off-axis modulation in a comparable setup. This indicates that on-axis operation adds a single rank-one component (the unmodulated light) and otherwise leaves the channel structure essentially unchanged. The effective rank is set by the noise floor of the calibration rather than by the number of controlled channels: because the spectrum decays smoothly, lower-noise calibration, with higher camera bit depth or averaging of repeated measurements, would raise the number of channels resolved above it.

These observations motivate a design principle for MMF wavefront shaping: the modulator needs about as many independent channels as the fiber modes. Rather than maximizing the number of independently controllable pixels, the modulator should provide sufficient control of the accessible modal dimensionality of the fiber while preserving a high refresh rate. This design approach balances focusing fidelity with calibration and computational overhead. Furthermore, direct phase modulation simplifies pattern generation by avoiding off-axis holographic encoding, both when preparing the input patterns and when generating the optimized focusing patterns. For the fiber used here, the principle calls for roughly three and a half times the channels controlled in this experiment, which can be achieved by future larger arrays or tiling of several devices. Because applications are limited by both photon budget and update rate, the product $\mathcal{M} = \varepsilon f_r$ complements enhancement as a figure of merit when systems control a comparable number of channels. Under that condition the DPM exceeds the alternatives by at least two orders of magnitude (Fig. 2a).

We characterized a short, static fiber so that efficiency, speed and fidelity could each be measured under controlled conditions. Like other transmission-matrix methods, it relies on calibration and is sensitive to fiber perturbations [51]. The demonstrated high refresh rate and reduced pattern-generation overhead provide an opportunity for faster adaptive wavefront correction when combined with sufficiently rapid detection and low-latency feedback, for instance with a high-speed camera on a small region of interest or single detector and compressive calibration [41]. As a result, the approach could enable real-time compensation of fiber perturbations while improving imaging robustness in practical endomicroscopic environments.

The demonstrated combination of high-speed phase modulation, high modulator efficiency, and thousands of controlled channels establishes a new operating regime for adaptive MMF endomicroscopy. These capabilities open opportunities for real-time volumetric imaging, dynamic wavefront correction, and MMF bending compensation, enabling extensions of minimally invasive imaging to changing biological environments.

## Methods

### Experimental setup

A 488 nm fiber-coupled laser is reshaped into an 8 x 1 mm rectangular beam to illuminate the active area of the DPM. The DPM operates in an on-axis configuration, with a polarizing beam splitter and quarter-wave plate separating the incident and modulated beams. The modulated beam is circularly polarized and relayed through a 4f system to an 0.65 numerical aperture (NA) objective, coupling the modulated light into a 34 mm-long MMF with a 100 μm core diameter and an NA of 0.37.

For TM calibration and focusing characterization, the distal fiber output is collected with an NA 0.4 objective and imaged with a tube lens onto an 8-bit CMOS camera synchronized with the DPM through hardware triggering. In the performed experiments, the 800 Hz camera frame rate limited TM acquisition to approximately 15 s. A pickoff mirror after the second lens allows relaying the focus through another 4f system with magnification optimized such that the speckle grain is the same size as the APD active area to measure the temporal response of the foci.

For fluorescence imaging, emission is collected through the same MMF, spectrally filtered, and relayed through a larger-core fiber to a Hamamatsu PMT. The amplified signal is digitized by a high-speed DAQ module, hardware-triggered by the DPM to synchronize acquisition with the focal scan.

### Noise floor estimation

For the non-square *m*-by-*n* shape of the TM, we define the aspect ratio

$$\beta = \frac{n}{m} \tag{7}$$

Under the assumption that the measurement noise is independent, identically distributed, and approximately spatially uncorrelated [5, 59], the MP model predicts that the singular values of the noise component lie approximately within

$$[\hat{a}(1 - \sqrt{\beta}), \hat{a}(1 + \sqrt{\beta})] \tag{8}$$

where $\hat{a}$ is the estimated noise scale from the spectral tail. Rather than estimating the noise scale from the complete singular-value spectrum using the Gavish-Donoho formulation [59], we use a tail-selection search algorithm to identify the portion of the spectrum that best fits the MP distribution. For each candidate, the remaining singular values are rescaled by matching their median to the median of the corresponding MP distribution, and the concordance between their cumulative distribution functions (CDF) is quantified using the Kolmogorov-Smirnov distance [63, 64]. A penalty discourages selecting a very short tail, while the lowest penalized discrepancy determines the selected tail and its estimated noise scale.

The noise threshold is then calculated as

$$\tau = \lambda(\beta)\hat{a} \tag{9}$$

where $\lambda(\beta)$ is the aspect ratio hard-threshold coefficient. Therefore, singular values above $\tau$ are considered the number of eigenchannels accessible in the fiber, and a total of 2,649 channels are being addressed. The unmodulated light not only reduces the expected enhancement but also increases noise during TM measurement, lowering the signal-to-noise ratio (SNR) (uses up part of the dynamic range and adds shot noise).

We independently estimate the noise from the spatial-frequency content of the measured TM output fields. The measured fields have a finite spatial-frequency support, and frequencies beyond this support provide an estimate of the measurement noise. Using a margin of 1-5% beyond the signal-support cutoff gives a retained rank of approximately 2,610, within 1.5% of the spectral-tail estimate calculated above.

**Image reconstruction**

The fluorescence image reconstructed through the fiber is a single-pixel measurement obtained by sequentially scanning the focus across the sample. Each measured value corresponds to the total fluorescence generated by the overlap between the fluorophore distribution and the spatially varying PSF at that scan position. Because the detected signal contains contributions from both in-focus and out-of-focus fluorescence, the directly reconstructed image exhibits reduced contrast because fluorescence excited by the speckle background adds to each measurement. Other reconstructions algorithms have been proposed for resolution enhancement in MMF imaging [65,66].

To improve the reconstruction fidelity of the fluorescence distribution, we performed spatially variant Richardson-Lucy (RL) deblurring using the experimentally measured high dynamic range PSF, obtained by combining low- and high-exposure measurements [61, 62]. The complete set of PSFs defines the system matrix $M$, which accounts for variations in the excitation focus. Starting from an initial estimate $s^{(0)}$, the sample is iteratively updated according to

$$s^{(k+1)} = s^k \odot M^T \left[\frac{r}{Ms^k}\right] \tag{10}$$

where $s^k$ is the estimated fluorophore distribution at iteration $k$, $r$ is the measured fluorescence signal, $M$ is the spatially variant forward operator, $M^T$ is its adjoint. Multiplication and division are performed elementwise.

## Data availability

Data underlying the results presented in this paper may be obtained from the corresponding author upon reasonable request.

## Acknowledgements

We acknowledge Lars Eng for his support of this work through Silicon Light Machines. We thank Eyal Niv for useful discussions on FPGA and software troubleshooting.

## Autor contributions

R.P. conceived and supervised the project. S.H. and A.P. designed, fabricated, and characterized the DPM. P.R.R. built the optical system, developed the DPM control software, performed the experiments and simulations, and analyzed the data, with supervision from A.M.C.A., O.T., and RP. P.R.R. and R.P. wrote the manuscript with input from all authors.

## Funding

This work was supported by the U.S. National Science Foundation awards 1548924 and 2415645.

## Competing interests

S.H. and A.P. are employed by Silicon Light Machines, which develops the DPM; the device is not yet commercially available. R.P. is the founder of Modendo Inc., and A.M.C.A. and O.T. are employees of Modendo Inc., a company developing multimode-fiber endomicroscopy.

## References

1. Snyder, A. W. & Love, J. D. *Optical Waveguide Theory* (Chapman and Hall, 1983).
2. Goodman, J. W. *Speckle Phenomena in Optics: Theory and Applications* (Roberts and Company Publishers, 2007).
3. Popoff, S. M. et al. Measuring the transmission matrix in optics: an approach to the study and control of light propagation in disordered media. *Phys. Rev. Lett.* **104**, 100601 (2010).
4. Popoff, S., Lerosey, G., Fink, M. *et al.* Image transmission through an opaque material. *Nat Commun* **1**, 81 (2010).
5. Popoff, S., Lerosey, G., Fink, M., Boccara, A. C. & Gigan, S. Controlling light through optical disordered media: transmission matrix approach. *New J. Phys.* **13**, 123021 (2011).
6. Čižmár, T. & Dholakia, K. Shaping the light transmission through a multimode optical fibre: complex transformation analysis and applications in biophotonics. *Opt. Express* **19**, 18871–18884 (2011).
7. Vellekoop, I. M. & Mosk, A. P. Focusing coherent light through opaque strongly scattering media. Opt. Lett. **32**, 2309–2311 (2007).
8. Yaqoob, Z., Psaltis, D., Feld, M. S. & Yang, C. Optical phase conjugation for turbidity suppression in biological samples. *Nat. Photon.* **2**, 110–115 (2008).
9. Vellekoop, I. M., Lagendijk, A. & Mosk, A. P. Exploiting disorder for perfect focusing. *Nat. Photon*. **4**, 320–322 (2010).
10. Akbulut, D., Huisman, T. J., van Putten, E. G., Vos, W. L. & Mosk, A. P. Focusing light through random photonic media by binary amplitude modulation. *Opt. Express* **19**, 4017-4029 (2011)
11. Cao, H., Mosk, A.P. & Rotter, S. Shaping the propagation of light in complex media. *Nat. Phys.* **18**, 994–1007 (2022).
12. Mosk, A.P., Lagendijk, A., Lerosey, G. *et al.* Controlling waves in space and time for imaging and focusing in complex media. *Nature Photon.* **6**, 283–292 (2012).
13. Guan, Y., Katz, O., Small, E., Zhou, J. & Silberberg, Y. Polarization control of multiply scattered light through random media by wavefront shaping. *Opt. Lett*. **37**, 4663–4665 (2012).
14. Xiong, W., Hsu, C.W., Bromberg, Y. *et al.* Complete polarization control in multimode fibers with polarization and mode coupling. *Light Sci Appl* **7**, 54 (2018).
15. McCabe, D. J. et al. Spatio-temporal focusing of an ultrafast pulse through a multiply scattering medium. *Nat. Commun.* **2**, 447 (2011).
16. Katz, O., Small, E., Bromberg, Y. & Silberberg, Y. Focusing and compression of ultrashort pulses through scattering media. *Nat. Photon.* **5**, 372–377 (2011).
17. Bender, N., Haig, H., Christodoulides, D., & Wise, F. W. Spatial, spectral, and temporal light control in disordered multimode fiber by spatial phase modulation. *APL Photon.* **11,** 026110 (2026).
18. Choi, Y. et al. Measurement of the time-resolved reflection matrix for enhancing light energy delivery into a scattering medium. *Phys. Rev. Lett.* **111**, 243901 (2013).
19. Aulbach, J., Gjonaj, B., Johnson, P. M., Mosk, A. P. & Lagendijk, A. Control of light transmission through opaque scattering media in space and time. *Phys. Rev. Lett.* **106**, 103901 (2011).
20. Tzang, O., Caravaca-Aguirre, A.M., Wagner, K. & Piestun, R. Adaptive wavefront shaping for controlling nonlinear multimode interactions in optical fibres. *Nature Photon* **12**, 368–374 (2018).
21. Bender, N., Yılmaz, H., Bromberg, Y., and Cao, H. Customizing speckle intensity statistics. *Optica* **5**, 595-600 (2018).
22. Devaud, L. et al. Speckle Engineering Through Singular Value Decomposition of the Transmission Matrix. *Phys. Rev. Lett.* **127**, 093903 (2021).
23. Riera Rodríguez, P. & Piestun, R. Random speckle patterns with azimuthal propagation order. *Photon. Res*. **14**, 4609-4618 (2026).
24. Čižmár, T. & Dholakia, K. Exploiting multimode waveguides for pure fibre-based imaging. *Nat Commun* **3**, 1027 (2012).

25. Papadopoulos, I. N., Farahi, S., Moser, C. & Psaltis, D. High-resolution, lensless endoscope based on digital scanning through a multimode optical fiber. *Biomed. Opt. Express* **4**, 260-270 (2013).
26. Caravaca-Aguirre, A. M. & Piestun, R. Single multimode fiber endoscope. *Opt. Express* **25**, 1656-1665 (2017).
27. Loterie, D., Psaltis, D. & Moser, C. Confocal microscopy via multimode fibers: fluorescence bandwidth. *Proc SPIE* **9717**, 97171C (2016).
28. Plöschner, M., Kollárová, V., Dostál, Z. *et al.* Multimode fibre: Light-sheet microscopy at the tip of a needle. *Sci Rep* **5**, 18050 (2015).
29. Singh, S., Labouesse, S. & Piestun, R. Multiview scattering scanning imaging confocal microscopy through a multimode fiber. *IEEE Trans. Comput Imaging* **9**, 159–171 (2023).
30. Li, Q. *et al*. Phase imaging through a single multimode fiber. *Opt. Lett.* **49**, 5823-5826 (2024).
31. Ivanina, A., Marshall, M., Abrashitova, K., van Leeuwen, T., & Amitonova, L. V. Quantitative phase imaging with a multimode fiber. *APL Photon.* **10**, 046116 (2025).
32. Morales-Delgado, E. E., Psaltis, D. & Moser, C. Two-photon imaging through a multimode fiber. *Opt. Express* **23**, 32158-32170 (2015).
33. Deng, S., Loterie, D., Konstantinou, G., Psaltis, D., & Moser, C. Raman imaging through multimode sapphire fiber. *Opt. Express* **27**, 1090-1098 (2019).
34. Pikálek, T., Stibůrek, M., Simpson, S., Čižmár, T., & Trägårdh, J. Suppression of the non-linear background in a multimode fibre CARS endoscope. *Biomed. Opt. Express* **13**, 862-874 (2022).
35. Trägårdh, J. *et al*. Label-free CARS microscopy through a multimode fiber endoscope. *Opt. Express* **27**, 30055-30066 (2019).
36. Sivankutty, S. *et al*. Ultra-thin rigid endoscope: two-photon imaging through a graded-index multi-mode fiber. *Opt. Express* **24**, 825-841 (2016).
37. Gomes, A. *et al*. Funnelling super-Resolution STED Microscopy through Multimode Fibres. Preprint at https://doi.org/10.48550/arXiv.2501.13572 (2025).
38. Zhao, T., Zhang, M., Ourselin, S. & Xia, W. Wavefront Shaping-Assisted Forward-Viewing Photoacoustic Endomicroscopy Based on a Transparent Ultrasound Sensor. *Appl. Sci.* **12**, 12619 (2022).
39. Zhao, T. *et al*. Ultrathin, high-speed, all-optical photoacoustic endomicroscopy probe for guiding minimally invasive surgery. *Biomed. Opt. Express* **13**, 4414-4428 (2022).
40. Mezil, S. *et al*. Single-shot hybrid photoacoustic-fluorescent microendoscopy through a multimode fiber with wavefront shaping. *Biomed. Opt. Express* **11**, 5717-5727 (2020).
41. Caravaca-Aguirre, A. M. *et al*. Hybrid photoacoustic-fluorescence microendoscopy through a multimode fiber using speckle illumination. *APL Photon.* **4**, 096103 (2019).
42. Zhao, T., Ma, M. T., Ourselin, S., Vercauteren, T. & Xia, W. Video-rate dual-modal photoacoustic and fluorescence imaging through a multimode fibre towards forward-viewing endomicroscopy. *Photoacoustics* **25**, 100323 (2022).
43. Ohayon, S., Caravaca-Aguirre, A. M., Piestun, R., & Dicarlo, J. J. Minimally invasive multimode optical fiber microendoscope for deep brain fluorescence imaging. *Biomed. Opt. Express* **9**, 1492–1509 (2018).
44. Pikalek, T. *et al*. Hair-thin confocal fluorescence endo-microscopy for deep-brain in-vivo imaging. Preprint as https: //arxiv.org/abs/2512.19419. 2512.19419 (2025).
45. Du, Y. *et al*. Advancing the path to *in-vivo* imaging in freely moving mice via multimode-multicore fiber based holographic endoscopy. *Neurophotonics* **11**, S11506 (2024).
46. Stibůrek, M. *et al.* 110 μm thin endo-microscope for deep-brain in vivo observations of neuronal connectivity, activity and blood flow dynamics. *Nat Commun* **14**, 1897 (2023).
47. Wen, Z. *et al.* Single multimode fibre for in vivo light-field-encoded endoscopic imaging. *Nat. Photon.* **17**, 679–687 (2023).
48. Turtaev, S. *et al.* High-fidelity multimode fibre-based endoscopy for deep brain in vivo imaging. *Light Sci Appl* **7**, 92 (2018).

49. Bianchi, S. & Di Leonardo, R. A multi-mode fiber probe for holographic micromanipulation and microscopy. *Lab Chip* **12**, 635–639 (2012).
50. Conkey, D. B., Caravaca-Aguirre, A. M. & Piestun, R. High-speed scattering medium characterization with application to focusing light through turbid media. *Opt. Express* **20**, 1733-1740 (2012).
51. Caravaca-Aguirre, A. M., Niv, E., Conkey, D. B. & Piestun, R. Real-time resilient focusing through a bending multimode fiber. *Opt. Express* **21**, 12881-12887 (2013).
52. Tzang, O., Niv, E., Singh, S., Labouesse, S. & Piestun, R. Wavefront shaping in complex media with a 350 kHz modulator via a 1D-to-2D transform. *Nat. Photonics* **13**, 788–793 (2019).
53. Rocha, J. C. A. *et al.* Fast and light-efficient wavefront shaping with a MEMS phase-only light modulator. *Opt. Express* **32**, 43300-43314 (2024).
54. Turtaev, S. *et al.* Comparison of nematic liquid-crystal and DMD based spatial light modulation in complex photonics. *Opt. Express* **25**, 29874-29884 (2017).
55. Gomes, A. D., Turtaev, S., Du, Y., & Čižmár, T. Near perfect focusing through multimode fibres. *Opt. Express* **30**, 10645-10663 (2022).
56. Payne, A. P. et al. Planar Light Valve and Displacement Phase Modulator: MEMS-based spatial light modulators for high-speed amplitude and phase modulation. *J. Opt. Microsyst.* **6**, 014001 (2026).
57. Yoon, C. et al. Experimental measurement of the number of modes for a multimode optical fiber. *Opt. Lett.* **37**, 4558-4560 (2012).
58. Marchenko, V. A. & Pastur, L.A. Distribution of eigenvalues for some sets of random matrices. *Math. USSR-Sb.* **1**, 457-483 (1967).
59. Gavish, M. & Donoho, D. L. The Optimal Hard Threshold for Singular Values is $4/\sqrt{3}$**.** *IEEE Trans. Inf. Theory* **60**, 5040-5053 (2014).
60. Ammar, A., Şener, S. F., Ercan, M. & Yilmaz, H. Upper bounds on focusing light through multimode fibers. *Phys. Rev. Appl*. **24**, 054006 (2025).
61. Richardson, W.H. Bayesian-Based Iterative Method of Image Reconstruction. *J. Opt. Soc. Am.* **62**, 55-59 (1972).
62. Lucy, L.B. An iterative technique for the rectification of observed distributions. *Astron.* **79**, 745-754 (1974).
63. Kolmogorov, A. Sulla determinazione empirica di una leffe di distribuzione. *G. Ist. Ital. Attuari.* **4**, 83-97 (1933).
64. Smirnov, N. Table for Estimating the Goodness of Fit of Empirical Distributions *Ann. Math. Statist.* **19**, 279-281 (1948).
65. Tučková, T. *et al.* Computational image enhancement of multimode fibre-based holographic endo-microscopy: harnessing the muddy modes, *Opt. Express* **29**, 38206-38220 (2021).
66. Turcotte, R. *et al.* Deconvolution for multimode fiber imaging: modeling of spatially variant PSF, *Biomed. Opt. Express* **11**, 4759-4771 (2020).